\documentclass[journal]{IEEEtran}
\IEEEoverridecommandlockouts
\usepackage{lineno}
\usepackage{hyperref}
\usepackage{cite}
\usepackage{amsmath,amssymb,amsfonts,cases} 
\usepackage{amsmath}
\usepackage{amsthm}
\DeclareMathOperator*{\argmax}{argmax}

\usepackage{graphicx}
\usepackage{textcomp}
\usepackage{xcolor}
\usepackage{graphicx}
\usepackage{float}
\usepackage{subfigure}
\usepackage{amsmath,mleftright}
\usepackage{amsfonts,amssymb}
\usepackage{mathrsfs}
\usepackage{mathtools}
\usepackage{algorithm}
\usepackage{algorithmic}
\usepackage{bm}
\usepackage{multirow}
\usepackage{array}
\usepackage{amssymb}
\usepackage{amsmath}
\usepackage{cite}
\usepackage{url}
\usepackage{xcolor}
\usepackage{cite,graphicx,amsmath,amssymb}
\usepackage{subfigure}
\usepackage{fancyhdr}
\usepackage{mdwmath}
\usepackage{mdwtab}
\usepackage{caption}
\usepackage{amsthm}
\usepackage{setspace}
\usepackage{bm}
\usepackage{mathtools}
\usepackage{dsfont}
\usepackage{bbm}
\usepackage{framed}

\newtheorem{theorem}{Theorem}

\newtheorem{lemma}{Lemma}

\newtheorem{corollary}{Corollary}

\allowdisplaybreaks[4]
\makeatletter
\newcommand{\biggg}{\bBigg@{3}}
\newcommand{\Biggg}{\bBigg@{3.5}}
\makeatother
\makeatletter
\renewcommand{\maketag@@@}[1]{\hbox{\m@th\normalsize\normalfont#1}}%
\makeatother
\def\BibTeX{{\rm B\kern-.05em{\sc i\kern-.025em b}\kern-.08em
    T\kern-.1667em\lower.7ex\hbox{E}\kern-.125emX}}
    \expandafter\def\expandafter\normalsize\expandafter{%
    \normalsize%
    \setlength\abovedisplayskip{4pt}%
    \setlength\belowdisplayskip{4pt}%
    \setlength\abovedisplayshortskip{2pt}%
    \setlength\belowdisplayshortskip{2pt}%
}

\begin{document}
\title{Fast Pinching-Antenna Activation for AirComp}
\author{Zhenqiao Cheng, Boqun Zhao, Chongjun Ouyang, and Xingqi Zhang\vspace{-10pt}
\thanks{Z. Cheng is with the 6G Research Centre, China Telecom Beijing Research Institute, Beijing 102209, China (e-mail: zhenqiao.cheng@engineer.com).}
\thanks{B. Zhao and X. Zhang are with the Department of Electrical and Computer Engineering, University of Alberta, Edmonton, AB T6G 2R3, Canada (e-mail: \{boqun1, xingqi.zhang\}@ualberta.ca).}
\thanks{C. Ouyang is with the School of Electronic Engineering and Computer Science, Queen Mary University of London, London E1 4NS, U.K. (e-mail: c.ouyang@qmul.ac.uk).}}
\maketitle

\begin{abstract}
A pinching-antenna system (PASS) is considered for over-the-air computation (AirComp). Multiple dielectric waveguides are deployed at the base station, and one pinching antenna (PA) is activated on each waveguide. For practical implementation, each PA is restricted to a finite set of preconfigured locations. The resulting discrete activation problem is formulated to minimize the AirComp mean-squared error (MSE). After the optimal aggregation vector is derived, the minimum MSE is expressed through an inverse Gram matrix. A rank-one recursion is derived to evaluate the exact MSE reduction produced by each candidate. Greedy search and beam search are then developed for fast tree search. To further reduce complexity, coherent aggregation search is proposed from a first-order MSE approximation. It admits a separable closed-form selection rule and is asymptotically optimal in the low signal-to-noise ratio (SNR) regime. Numerical results show that the proposed methods substantially improve the AirComp accuracy of PASS over conventional antenna arrays.
\end{abstract}


\section{Introduction}
Over-the-air computation (AirComp) exploits the waveform-superposition property of a multiple-access channel to aggregate distributed data directly over the air \cite{goldenbaum2013harnessing,zhu2019mimoaircomp}. It can therefore support low-latency sensing and edge-learning applications without separately decoding every user \cite{yang2020federated}. The computation accuracy is commonly measured by the mean-squared error (MSE), which depends critically on the alignment between the user channels and the receive aggregation vector.

Reconfigurable antenna technologies have been introduced to improve this alignment by changing the wireless channel seen by the receiver. Typical examples include reconfigurable intelligent surfaces \cite{fang2021over}, fluid antennas \cite{zhang2024fluid}, and movable antennas \cite{cheng2023movable,li2024over}, among others. However, many existing architectures mainly reshape wavelength-scale channel responses and are less effective in reducing large-scale path loss and signal blockage. Pinching-antenna systems (PASS) provide a new solution by activating pinching antennas (PAs) at flexible positions along dielectric waveguides \cite{suzuki2022pinching,ding2024flexible}. Since the receive locations can be moved closer to the users or around blocked regions, PASS can reduce large-scale path loss and alleviate signal blockage \cite{liu2025pinching,liu2026pinchingt,liu2026survey}.

PASS-aided AirComp has recently been studied through joint PA positioning, power control, and receive combining \cite{lyu2025aircomp}. Segmented-waveguide architectures have also been proposed for signal aggregation \cite{gu2026segmented}. These studies usually treat PA positions as continuous variables and solve the resulting problem by iterative optimization. In practical PASS hardware, however, mechanical, magnetic, or electronic control mechanisms may support only a finite number of activation positions \cite{liu2026pinchingt,liu2026survey}. This hardware constraint motivates a discrete model in which each waveguide activates one PA from a predefined candidate set \cite{wang2025antenna,Cheng2026optimal}. The resulting search space is combinatorial and naturally admits a tree representation.

This letter studies fast discrete PA activation for an uplink multiuser PASS with one active PA on each waveguide. The contributions are summarized as follows. First, we derive the optimal aggregation vector, formulate the resulting minimum MSE through an inverse Gram matrix, and obtain a rank-one recursion for the exact MSE reduction of each candidate PA. Second, we cast discrete PA activation as a multi-layer tree search and propose two search algorithms: greedy search (GS), which retains only the best partial path at each layer, and beam search (BeS), which retains several promising paths to balance complexity and performance. Third, we propose coherent aggregation search (CAS) from a first-order MSE approximation. CAS yields a separable closed-form rule, reveals the roles of in-waveguide attenuation, free-space path loss, and coherent phase alignment, and is asymptotically optimal in the low signal-to-noise ratio (SNR) regime. Numerical results show that CAS, GS, and BeS achieve distinct complexity-performance tradeoffs and substantially reduce the MSE compared with the conventional fixed array.

\begin{figure}[!t]
\centering
\includegraphics[width=0.45\textwidth]{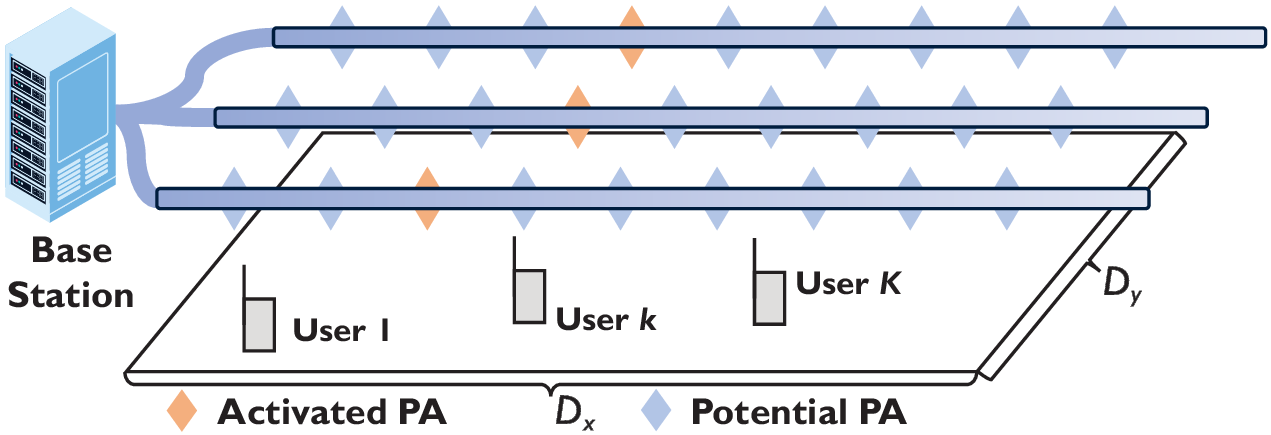}
\caption{Illustration of an uplink multiuser PASS for AirComp.}
\label{fig:system}
\vspace{-10pt}
\end{figure}

\section{System Model}
Consider the PASS-enabled AirComp system illustrated in Fig.~\ref{fig:system}. A base station (BS) receives the simultaneous transmissions of $K$ single-antenna users through $M$ dielectric waveguides. The user set and waveguide set are denoted by ${\mathcal K}\triangleq\{1,\ldots,K\}$ and ${\mathcal M}\triangleq\{1,\ldots,M\}$, respectively. User $k$ is located at ${\mathbf u}_k=[x_k,y_k,0]^{\mathsf T}$. The waveguides are deployed along the $x$-axis at height $h$ and are separated along the $y$-axis. The feed point of waveguide $m$ is ${\bm\psi}_0^m=[\psi_{\rm w},\psi_0^m,h]^{\mathsf T}$.

One PA is activated on each waveguide \cite{zeng2025energy,zeng2025sum,ouyang2025uplink}. The available longitudinal coordinates form the candidate set ${\mathcal L}\triangleq\{\psi_1,\ldots,\psi_L\}$, and ${\mathcal L}_{\rm id}\triangleq\{1,\ldots,L\}$ is its index set. If candidate $\ell_m$ is selected on waveguide $m$, the PA position is ${\bm\psi}_{\ell_m}^m=[\psi_{\ell_m},\psi_0^m,h]^{\mathsf T}$. The activation vector is ${\bm\ell}=[\ell_1,\ldots,\ell_M]^{\mathsf T}$.

PASS is envisioned for operation in high-frequency bands \cite{suzuki2022pinching}, where line-of-sight (LoS) propagation dominates \cite{liu2023near-field,ouyang2024primer}. A free-space LoS channel model is therefore adopted. Under this model, the spatial channel coefficient from user $k$ to candidate $\ell$ on waveguide $m$ is modeled as \cite{liu2023near-field,ouyang2024primer}
\begin{align}
h_{\rm o}({\mathbf u}_k,{\bm\psi}_{\ell}^m)
=\frac{\sqrt{\eta}\,{\rm e}^{-{\rm j}k_0\|{\mathbf u}_k-{\bm\psi}_{\ell}^m\|}}
{\|{\mathbf u}_k-{\bm\psi}_{\ell}^m\|},
\end{align}
where $\eta\triangleq \frac{c^2}{16\pi^2f_{\rm c}^2}$, $c$ is the speed of light, $f_{\rm c}$ is the carrier frequency, $\lambda$ is the free-space wavelength, and $k_0=\frac{2\pi}{\lambda}$ is the wavenumber. The in-waveguide propagation coefficient from the activated PA to the feed point is modeled as \cite{pozar2021microwave}
\begin{align}
h_{\rm i}({\bm\psi}_{\ell}^m,{\bm\psi}_{0}^{m})
=10^{-\frac{\kappa}{20}\|{\bm\psi}_{\ell}^m-{\bm\psi}_{0}^{m}\|}
{\rm e}^{-{\rm j}\frac{2\pi}{\lambda_{\rm g}}
\|{\bm\psi}_{\ell}^m-{\bm\psi}_{0}^{m}\|},
\end{align}
where $\lambda_{\rm g}=\frac{\lambda}{n_{\rm eff}}$ is the guided wavelength and $n_{\rm eff}$ is the effective refractive index of the dielectric waveguide \cite{pozar2021microwave}. The parameter $\kappa$ denotes the average attenuation factor along the waveguide in dB/m \cite{yeh2008essence}. As a result, the end-to-end channel from user $k$ to the PA activated at waveguide $m$ is
\begin{align}
g_{m,k}(\ell)=h_{\rm i}({\bm\psi}_{\ell}^m,{\bm\psi}_{0}^{m})
h_{\rm o}({\mathbf u}_k,{\bm\psi}_{\ell}^m).
\end{align}

Let $s_k\sim{\mathcal{CN}}(0,1)$ denote the preprocessed datum of user $k$, with ${\mathbbmss E}[s_ks_{k'}^*]=0$ for $k\neq k'$. Following the classical
AirComp model \cite{yang2020federated}, the BS aims to recover the desired function, namely the sum
\begin{align}
s_{\rm sum}=\sum_{k=1}^{K}s_k={\mathbf 1}^{\mathsf T}{\mathbf s},
\end{align}
through the waveform superposition property of the wireless multiple-access channel, where ${\mathbf s}=[s_1,\ldots,s_K]^{\mathsf T}$. The users transmit with fixed powers collected in ${\mathbf P}={\rm diag}(P_1,\ldots,P_K)$. The received signal is
\begin{align}
{\mathbf y}={\mathbf G}({\bm \ell}){\mathbf P}^{\frac{1}{2}}{\mathbf s}+{\mathbf n},
\end{align}
where ${\mathbf G}({\bm \ell})\in{\mathbbmss C}^{M\times K}$ has entry $[{\mathbf G}({\bm \ell})]_{m,k}=g_{m,k}(\ell_m)$, and ${\mathbf{n}}\sim{\mathcal{CN}}({\mathbf{0}},\sigma^2{\mathbf{I}}_M)$ is the additive noise with covariance matrix $\sigma^2{\mathbf I}_M$. For an aggregation vector ${\mathbf v}\in{\mathbbmss C}^{M}$, the estimate is $\hat{s}_{\rm sum}={\mathbf v}^{\mathsf H}{\mathbf y}$. Its MSE is
\begin{align}
{\mathcal E}({\bm\ell},{\mathbf v})
&=\mathbbmss E[|\hat{s}_{\rm sum}-s_{\rm sum}|^2] =\mathbbmss E[|{\mathbf v}^{\mathsf H}{\mathbf y}-{\mathbf 1}^{\mathsf T}{\mathbf s}|^2]\\
&={\sigma^2}(\|{\mathbf H}^{\mathsf H}({\bm\ell}){\mathbf v}-\sigma^{-1}{\mathbf 1}\|_2^2
+\|{\mathbf v}\|_2^2),
\label{eq:mse_v}
\end{align}
where the $m$th row of ${\mathbf H}({\bm \ell})=\frac{1}{\sigma}{\mathbf G}({\bm \ell}){\mathbf P}^{\frac12}$ is
\begin{align}
{\mathbf h}_{m,\ell_m}^{\mathsf H}
=\frac{1}{\sigma}[g_{m,1}(\ell_m)\sqrt{P_1},\ldots,
g_{m,K}(\ell_m)\sqrt{P_K}].
\end{align}

For notational brevity, the dependence of ${\mathbf H}$ on ${\bm\ell}$ is omitted in the following derivation. Expanding \eqref{eq:mse_v} gives
\begin{align}
{\mathcal E}({\bm\ell},{\mathbf v})
&=K-2\sigma\Re\{{\mathbf v}^{\mathsf H}{\mathbf H}{\mathbf 1}\}
+{\sigma^2}{\mathbf v}^{\mathsf H}({\mathbf H}{\mathbf H}^{\mathsf H}
+{\mathbf I}_M){\mathbf v}.
\label{eq:mse_expand}
\end{align}
Setting the derivative of \eqref{eq:mse_expand} with respect to ${\mathbf v}^*$ to zero yields the linear minimum mean-squared error (LMMSE) aggregation vector
\begin{align}
{\mathbf v}^{\star}({\bm\ell})
=\frac{1}{\sigma}({\mathbf H}({\bm\ell}){\mathbf H}^{\mathsf H}({\bm\ell})
+{\mathbf I}_M)^{-1}{\mathbf H}({\bm\ell}){\mathbf 1}.
\label{eq:v_opt}
\end{align}
Substituting \eqref{eq:v_opt} into \eqref{eq:mse_expand} yields
\begin{align}
{\mathcal E}^{\star}({\bm\ell})
&=K-{\mathbf 1}^{\mathsf H}{\mathbf H}^{\mathsf H}
({\mathbf H}{\mathbf H}^{\mathsf H}+{\mathbf I}_M)^{-1}
{\mathbf H}{\mathbf 1}.
\label{eq:min_mse_1}
\end{align}
By the Woodbury identity \cite{horn2012matrix},
\begin{align}
({\mathbf I}_K+{\mathbf H}^{\mathsf H}{\mathbf H})^{-1}
={\mathbf I}_K-{\mathbf H}^{\mathsf H}
({\mathbf I}_M+{\mathbf H}{\mathbf H}^{\mathsf H})^{-1}{\mathbf H}.
\label{eq:woodbury}
\end{align}
Combining \eqref{eq:min_mse_1} and \eqref{eq:woodbury}, the minimum MSE can be equivalently written as
\begin{align}
{\mathcal E}^{\star}({\bm\ell})
={\mathbf 1}^{\mathsf H}
({\mathbf I}_K+{\mathbf H}^{\mathsf H}({\bm\ell})
{\mathbf H}({\bm\ell}))^{-1}{\mathbf 1}.
\label{eq:min_mse}
\end{align}
The discrete PA activation problem is therefore
\begin{align}
\min_{{\bm\ell}}\quad&
{\mathcal E}^{\star}({\bm\ell}) \label{eq:problem}\\
{\rm s.t.}\quad&\ell_m\in{\mathcal L}_{\rm id},\quad m\in{\mathcal M}.
\nonumber
\end{align}
Problem \eqref{eq:problem} contains $L^M$ feasible activation vectors.

\section{Fast PA Activation}
\subsection{MSE-Reduction Recursion}
\begin{figure}[!t]
\centering
\includegraphics[width=0.25\textwidth]{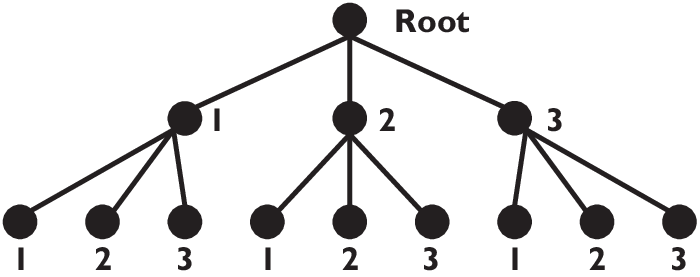}
\caption{Example of the search tree. $M=2$ and $L=3$.}
\label{fig:tree}
\vspace{-10pt}
\end{figure}

Problem \eqref{eq:problem} is represented by an $M$-layer tree, as exemplified in Fig.~\ref{fig:tree}. Layer $m$ selects one candidate on waveguide $m$, and every root-to-leaf path specifies one complete activation vector.

For a partial path of length $n$, define
\begin{align}
{\mathbf G}_n&\triangleq({\mathbf I}_K+{\mathbf H}_n^{\mathsf H}{\mathbf H}_n)^{-1},\\
{\mathcal E}_n&\triangleq{\mathbf 1}^{\mathsf H}{\mathbf G}_n{\mathbf 1},
\end{align}
where ${\mathbf H}_n$ contains the first $n$ selected rows. At the root, ${\mathbf G}_0={\mathbf I}_K$ and ${\mathcal E}_0=K$.

Suppose candidate $\ell$ on waveguide $m=n+1$ is appended to the path. The Sherman-Morrison identity \cite{horn2012matrix} yields
\begin{align}
{\mathbf G}_{n+1}
={\mathbf G}_n-
\frac{{\mathbf G}_n{\mathbf h}_{m,\ell}
{\mathbf h}_{m,\ell}^{\mathsf H}{\mathbf G}_n}
{1+{\mathbf h}_{m,\ell}^{\mathsf H}{\mathbf G}_n{\mathbf h}_{m,\ell}}.
\label{eq:G_update}
\end{align}
Consequently,
\begin{align}
{\mathcal E}_{n+1}
={\mathcal E}_n-\Delta_{m,\ell,n},
\label{eq:mse_recursion}
\end{align}
where the exact MSE reduction is
\begin{align}
\Delta_{m,\ell,n}
=\frac{|{\mathbf h}_{m,\ell}^{\mathsf H}{\mathbf G}_n{\mathbf 1}|^2}
{1+{\mathbf h}_{m,\ell}^{\mathsf H}{\mathbf G}_n{\mathbf h}_{m,\ell}}.
\label{eq:reduction}
\end{align}
Thus, every child node can be evaluated by one quadratic form and one rank-one update.

\subsection{Greedy and Beam Searches}
For general SNR regimes, the interaction among waveguides is captured by the exact recursion in \eqref{eq:reduction}. GS traverses the search tree with only one survivor path. At layer $m$, it selects
\begin{align}
\ell_m^{\rm GS}
=\argmax_{\ell\in{\mathcal L}_{\rm id}}
\Delta_{m,\ell,m-1}.
\label{eq:gs_rule}
\end{align}
This rule has a useful AirComp interpretation. Define the residual aggregation direction
\begin{align}
{\mathbf r}_n={\mathbf G}_n{\mathbf 1}.
\label{eq:residual_direction}
\end{align}
Initially, ${\mathbf r}_0={\mathbf 1}$, so all users contribute equally to the desired sum. After several PAs have been activated, the already observed user components are suppressed through ${\mathbf G}_n$. The vector ${\mathbf r}_n$ then identifies the part of the summation direction that remains insufficiently observed. Hence, GS does not simply choose the strongest PA. It selects the candidate whose channel is most aligned with the remaining aggregation error. The denominator in \eqref{eq:reduction} penalizes redundant or inefficient observations.

GS is simple, but it retains only one partial activation path at each layer. A locally good PA selection may remove a partial path that becomes better after later waveguides are activated. To alleviate this issue, BeS keeps $B$ promising partial paths after each layer, where $B$ is the beam width. At layer $m$, each surviving path from layer $m-1$ is expanded by all $L$ candidate locations. For every expanded child, the MSE reduction is computed by \eqref{eq:reduction}, and the updated MSE is obtained from \eqref{eq:mse_recursion}. Among all expanded paths, BeS retains the $B$ paths with the smallest partial MSE. When $B=1$, BeS reduces to GS. When $B$ is sufficiently large, BeS becomes exhaustive search (ES). Therefore, BeS provides a tunable complexity-performance tradeoff between GS and ES. The unified implementation of GS and BeS is summarized in Algorithm~\ref{alg:gs_bes}.

\begin{algorithm}[!t]
\caption{GS and BeS for Discrete PA Activation}
\label{alg:gs_bes}
\begin{algorithmic}[1]
\STATE Initialize ${\mathcal P}_0=\{(\emptyset,{\mathbf I}_K,K)\}$.
\FOR{$m=1,\ldots,M$}
\STATE Set the child list ${\mathcal C}=\emptyset$.
\FOR{each $({\bm\ell}_{1:m-1},{\mathbf G}_{m-1},{\mathcal E}_{m-1})
\in{\mathcal P}_{m-1}$}
\FOR{$\ell=1,\ldots,L$}
\STATE Compute $\Delta_{m,\ell,m-1}$ from \eqref{eq:reduction}.
\STATE Set ${\mathcal E}_m={\mathcal E}_{m-1}-\Delta_{m,\ell,m-1}$.
\STATE Update ${\mathbf G}_m$ from \eqref{eq:G_update}.
\STATE Add $([{\bm\ell}_{1:m-1}^{\mathsf T},\ell]^{\mathsf T},
{\mathbf G}_m,{\mathcal E}_m)$ to ${\mathcal C}$.
\ENDFOR
\ENDFOR
\STATE Retain the $\min\{B,|{\mathcal C}|\}$ entries with the smallest MSE to form ${\mathcal P}_m$.
\ENDFOR
\STATE Output the path in ${\mathcal P}_M$ with the smallest MSE.
\end{algorithmic}
\end{algorithm}

\subsection{Coherent Aggregation Search}
Although GS is efficient, it still requires recursive matrix updates. To obtain a simpler rule, the low-SNR behavior of \eqref{eq:problem} is examined. When the effective channel is weak, ${\mathbf H}^{\mathsf H}{\mathbf H}$ is small, and it follows that
\begin{align}
({\mathbf I}_K+{\mathbf H}^{\mathsf H}{\mathbf H})^{-1}
\approx {\mathbf I}_K-{\mathbf H}^{\mathsf H}{\mathbf H}.
\end{align}
Therefore, the minimum AirComp MSE can be approximated as
\begin{align}
{\mathcal E}^{\star}({\bm \ell})
&={\mathbf 1}^{\mathsf H}({\mathbf I}_K+
{\mathbf H}^{\mathsf H}({\bm \ell}){\mathbf H}({\bm \ell}))^{-1}{\mathbf 1}\\
&\approx K-{\mathbf 1}^{\mathsf H}{\mathbf H}^{\mathsf H}({\bm \ell})
{\mathbf H}({\bm \ell}){\mathbf 1}.
\label{eq:low_snr_mse}
\end{align}
Since
\begin{align}
{\mathbf 1}^{\mathsf H}{\mathbf H}^{\mathsf H}({\bm \ell})
{\mathbf H}({\bm \ell}){\mathbf 1}
=\sum_{m=1}^{M}|{\mathbf h}_{m,\ell_m}^{\mathsf H}{\mathbf 1}|^2,
\label{eq:low_snr_sep}
\end{align}
the first-order low-SNR activation problem is separable across waveguides. This motivates CAS with the rule
\begin{align}
\ell_m^{\rm CAS}
=\argmax_{\ell\in{\mathcal L}_{\rm id}}
|{\mathbf h}_{m,\ell}^{\mathsf H}{\mathbf 1}|^2,\quad m\in{\mathcal M}.
\label{eq:cas_general}
\end{align}
The name follows from the metric $|{\mathbf h}_{m,\ell}^{\mathsf H}{\mathbf 1}|^2$, which measures the coherent aggregation gain produced by a candidate PA. CAS does not require tree search or matrix inversion and can be applied independently on all waveguides. Although it is motivated by a low-SNR expansion, the rule can also be used as a low-complexity activation method at general SNRs.

Substituting the PASS channel model into \eqref{eq:cas_general}, we have
\begin{align}
{\mathbf h}_{m,\ell}^{\mathsf H}{\mathbf 1}
=\frac{1}{\sigma}\sum_{k=1}^{K}\sqrt{P_k}g_{m,k}(\ell),
\label{eq:h1_pass}
\end{align}
where $g_{m,k}(\ell)=h_{\rm i}({\bm\psi}_{\ell}^m,{\bm\psi}_{0}^{m})
h_{\rm o}({\mathbf u}_k,{\bm\psi}_{\ell}^m)$. Hence,
\begin{align}
\ell_m^{\rm CAS}
=\argmax_{\ell\in{\mathcal L}_{\rm id}}
\left|h_{\rm i}({\bm\psi}_{\ell}^m,{\bm\psi}_{0}^{m})
\sum_{k=1}^{K}\sqrt{P_k}
h_{\rm o}({\mathbf u}_k,{\bm\psi}_{\ell}^m)\right|^2.
\label{eq:cas_pass}
\end{align}
Using \eqref{eq:cas_pass} together with the LoS and in-waveguide models gives
\begin{align}
\ell_m^{\rm CAS}
=\argmax_{\ell\in{\mathcal L}_{\rm id}}\;&
10^{-\frac{\kappa}{10}\|{\bm\psi}_{\ell}^m-{\bm\psi}_{0}^{m}\|}
\nonumber\\
&\times\left|\sum_{k=1}^{K}\sqrt{P_k}
\frac{\sqrt{\eta}{\rm e}^{-{\rm j}k_0\|{\mathbf u}_k-{\bm\psi}_{\ell}^m\|}}
{\|{\mathbf u}_k-{\bm\psi}_{\ell}^m\|}\right|^2 .
\label{eq:cas_closed}
\end{align}
Equation \eqref{eq:cas_closed} shows that CAS balances three physical factors: in-waveguide attenuation, free-space path loss, and coherent phase alignment among users. Moreover, from \eqref{eq:low_snr_mse}-\eqref{eq:low_snr_sep}, CAS is asymptotically optimal as the transmit powers vanish, since it solves the first-order approximation of \eqref{eq:problem}. In the same regime, ${\mathbf G}_n\approx{\mathbf I}_K$ for all fixed $n$. The GS reduction in \eqref{eq:reduction} then reduces to $|{\mathbf h}_{m,\ell}^{\mathsf H}{\mathbf 1}|^2$ up to a common first-order scaling. Hence, GS and CAS select the same candidates asymptotically at low SNR.

\begin{algorithm}[!t]
\caption{CAS for Discrete PA Activation}
\label{alg:cas}
\begin{algorithmic}[1]
\FOR{$m=1,\ldots,M$}
\FOR{$\ell=1,\ldots,L$}
\STATE Compute $|{\mathbf h}_{m,\ell}^{\mathsf H}{\mathbf 1}|^2$.
\ENDFOR
\STATE Set $\ell_m^{\rm CAS}$ according to \eqref{eq:cas_general}.
\ENDFOR
\STATE Output ${\bm\ell}^{\rm CAS}=[\ell_1^{\rm CAS},\ldots,\ell_M^{\rm CAS}]^{\mathsf T}$.
\end{algorithmic}
\end{algorithm}

\subsection{Computational Complexity}
CAS evaluates $L$ scalar metrics on each waveguide and has complexity ${\mathcal O}(MLK)$. For GS and BeS, each child evaluation in \eqref{eq:reduction} and \eqref{eq:G_update} costs ${\mathcal O}(K^2)$. GS visits
\begin{align}
N_{\rm GS}=ML
\end{align}
children and therefore has complexity ${\mathcal O}(MLK^2)$. BeS visits
\begin{align}
N_{\rm BeS}\leq L+(M-1)BL
\end{align}
children, since it visits $L$ children in the first layer and at most $BL$ children in each subsequent layer. Its complexity is ${\mathcal O}(N_{\rm BeS}K^2)$ and its memory cost is ${\mathcal O}(BK^2)$. In contrast, ES visits
\begin{align}
N_{\rm ES}=\sum_{m=1}^{M}L^m=\frac{L^{M+1}-L}{L-1}
\end{align}
tree nodes. Therefore, CAS, GS, and BeS avoid the exponential scaling of ES. The beam width $B$ controls the tradeoff: a small $B$ gives low complexity, while a larger $B$ improves the chance of retaining near-optimal activation paths.

\begin{figure}[!t]
\centering
    \subfigure[${\mathcal E}^{\star}({\bm\ell})$ versus SNR.]
    {
        \includegraphics[height=0.17\textwidth]{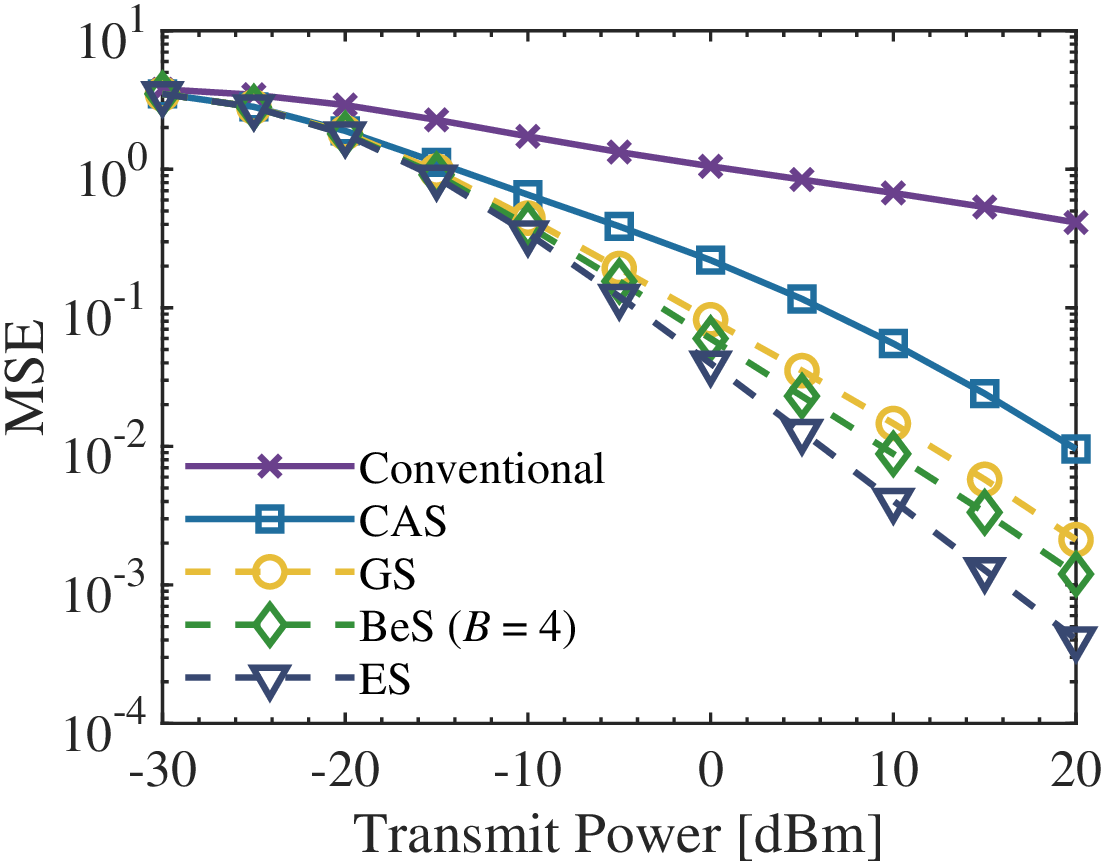}
	   \label{fig:fig1a_mse_snr_results}
    }
    \subfigure[${\mathcal E}^{\star}({\bm\ell})$ versus $D_x$.]
    {
        \includegraphics[height=0.17\textwidth]{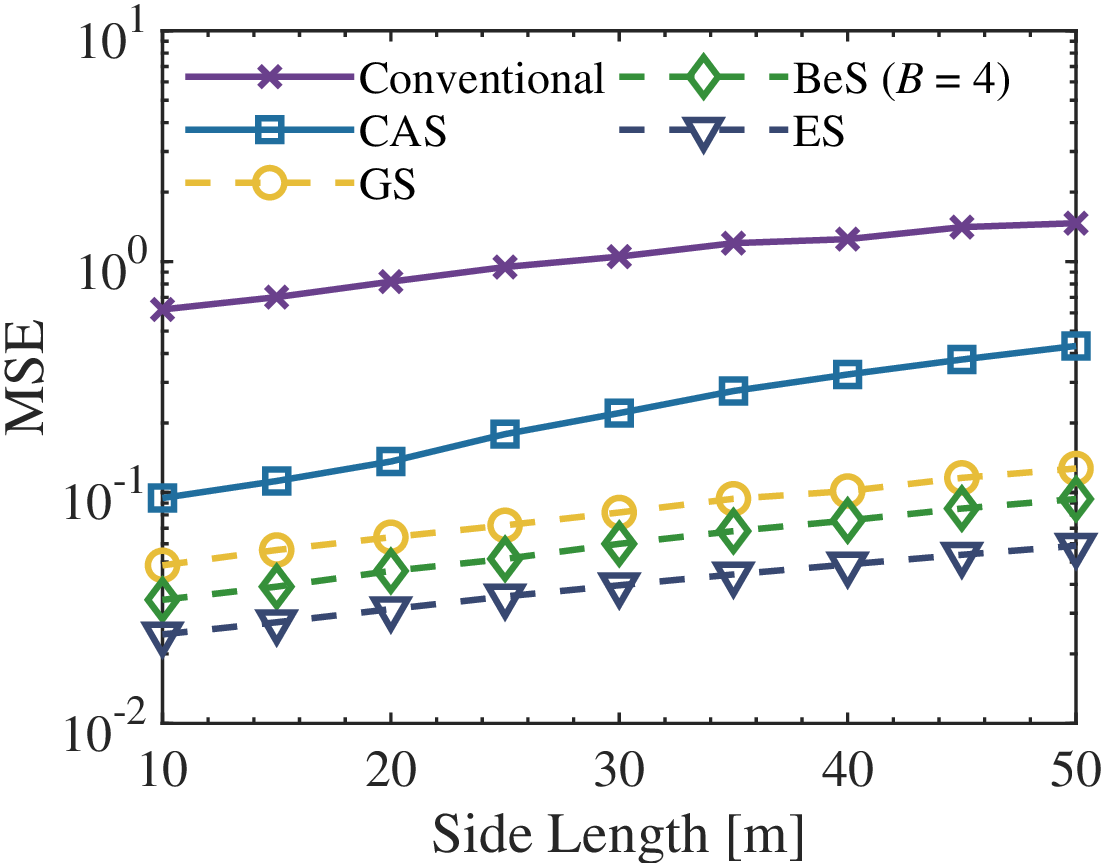}
	   \label{fig:fig1b_mse_Dx}
    }
\caption{AirComp MSE comparison.}
\label{fig:fig1_mse_results}
\vspace{-10pt}
\end{figure}

\begin{figure}[!t]
\centering
    \subfigure[MSE.]
    {
        \includegraphics[height=0.17\textwidth]{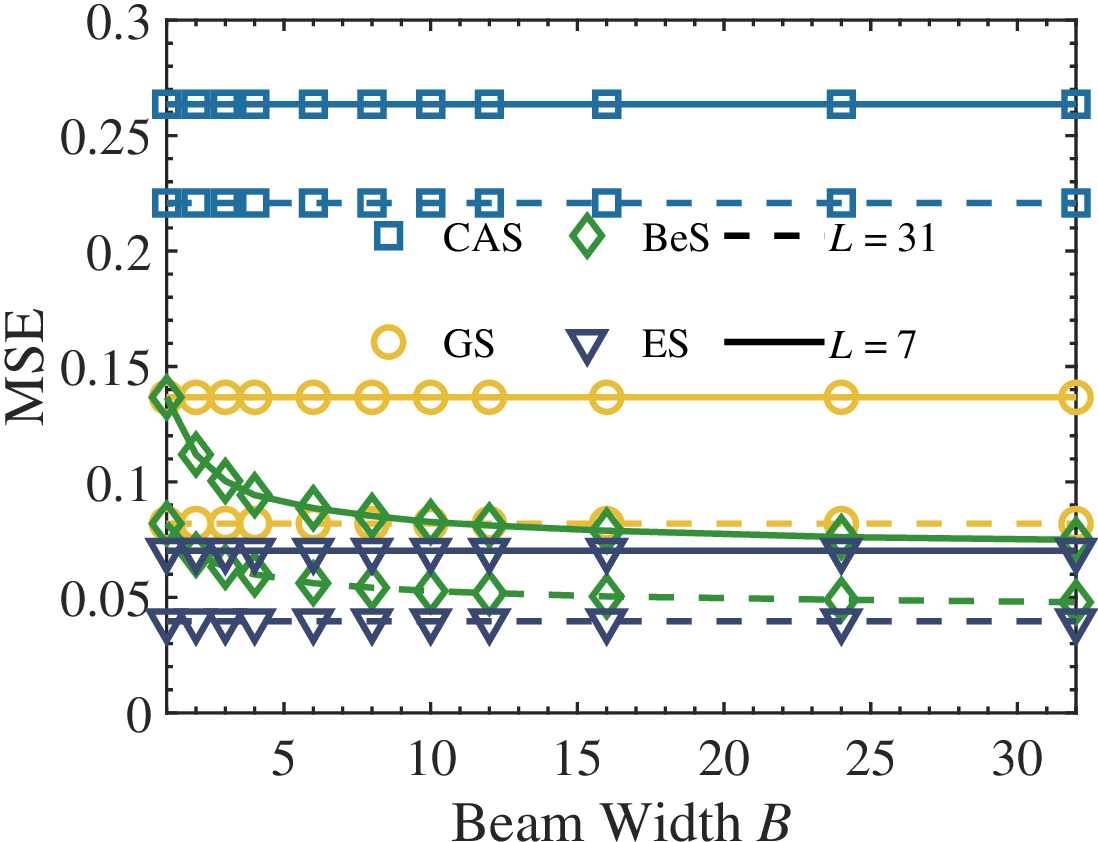}
	   \label{fig:fig2a_mse_beam_width}
    }
    \subfigure[Running time.]
    {
        \includegraphics[height=0.17\textwidth]{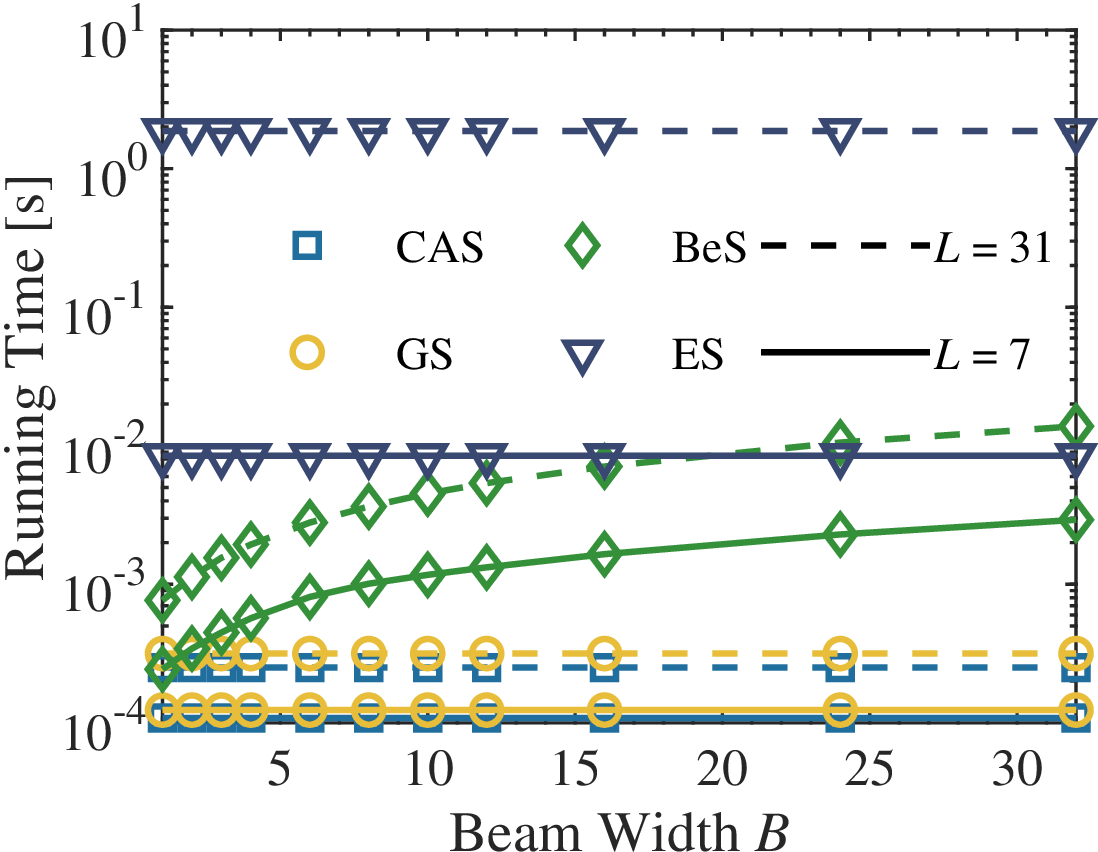}
	   \label{fig:fig2b_runtime_beam_width}
    }
\caption{Impact of beam width.}
\label{fig:fig_mse_beam_width}
\vspace{-10pt}
\end{figure}

\section{Numerical Results}
Numerical results are presented to validate the effectiveness and computational complexity of the proposed methods. Unless otherwise specified, we set $M=K=4$, $f_{\rm c}=28$ GHz, $n_{\rm eff}=1.4$, $\kappa=0.08$ dB/m, $h=3$ m, $P_1=\cdots=P_K=P=0$ dBm, and $\sigma^2=-90$ dBm. The users are independently and uniformly distributed over a $D_x\times D_y$ ground region with $D_x=30$ m and $D_y=10$ m. The waveguides span $[0,D_x]$ along the $x$-axis, with $\psi_0^m=-\frac{D_y}{2}+\frac{D_y(m-1)}{M-1}$ and $\psi_{\rm w}=0$. The $L$ candidate locations uniformly cover each waveguide, with spacing $\Delta=\frac{D_x}{L-1}$. The conventional benchmark is a centered $M$-element array parallel to the $y$-axis with half-wavelength spacing. All results are averaged over $1000$ independent user deployments. Running-time results are obtained using MATLAB R2025a on an Ubuntu server with two Intel Xeon Gold 5418Y CPUs, 96 logical CPUs, and 128 GB RAM.

Fig.~\ref{fig:fig1a_mse_snr_results} and Fig.~\ref{fig:fig1b_mse_Dx} compare the AirComp MSE versus SNR and the service-region length, respectively. PASS activation consistently outperforms the conventional fixed array because its receive locations are not confined to a centered aperture. Each PA can be activated at a favorable point on its waveguide, which shortens the average propagation distance, reduces large-scale path loss, avoids unfavorable fixed receive positions, and improves coherent aggregation. In Fig.~\ref{fig:fig1a_mse_snr_results}, CAS nearly coincides with GS and ES at low SNR, in agreement with the separable coherent aggregation metric in Section IV-C. At higher SNR, the interaction among selected PAs becomes more important, so GS improves over CAS through residual-error updates, while BeS approaches ES by retaining multiple promising paths. Fig.~\ref{fig:fig1b_mse_Dx} further shows that the PASS gain becomes more pronounced as $D_x$ increases. The conventional array suffers a larger distance penalty over a wider area, whereas PASS moves the activated PAs along the waveguides and offsets much of this degradation.

\begin{figure}[!t]
\centering
    \subfigure[MSE.]
    {
        \includegraphics[height=0.17\textwidth]{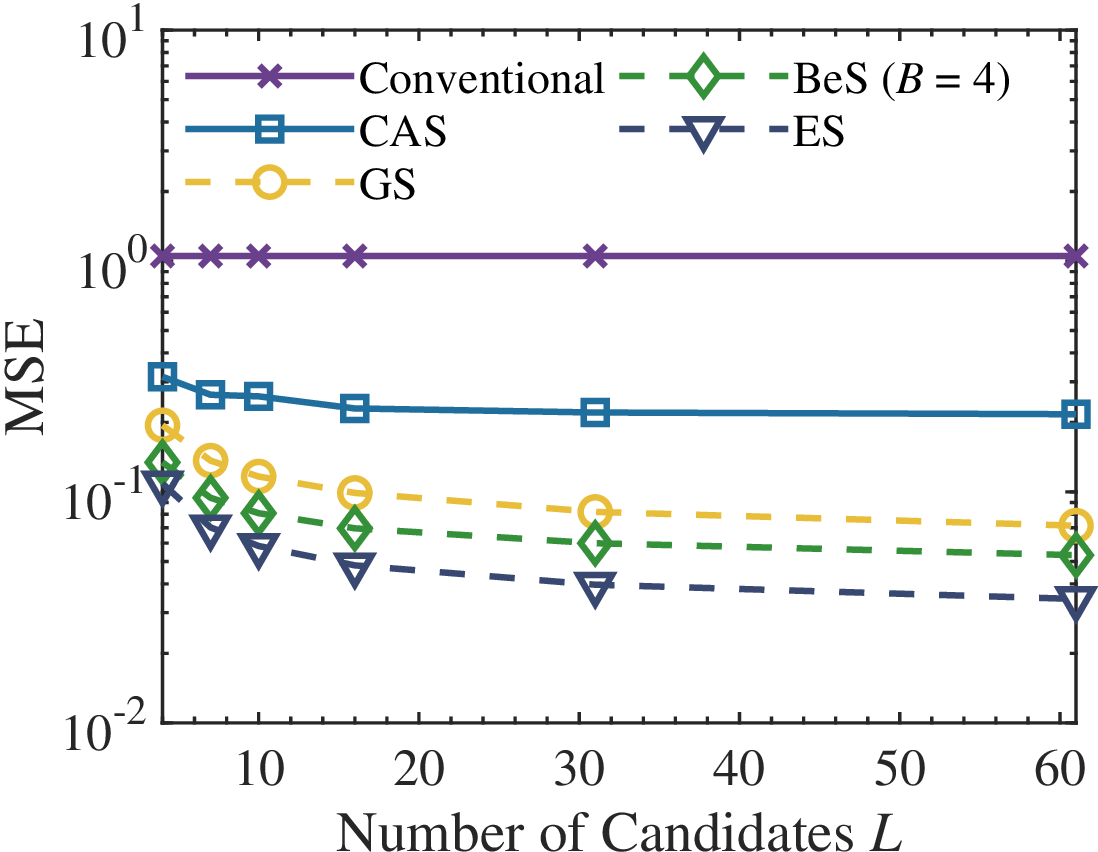}
	   \label{fig:fig3a_mse_L}
    }
    \subfigure[Running time.]
    {
        \includegraphics[height=0.17\textwidth]{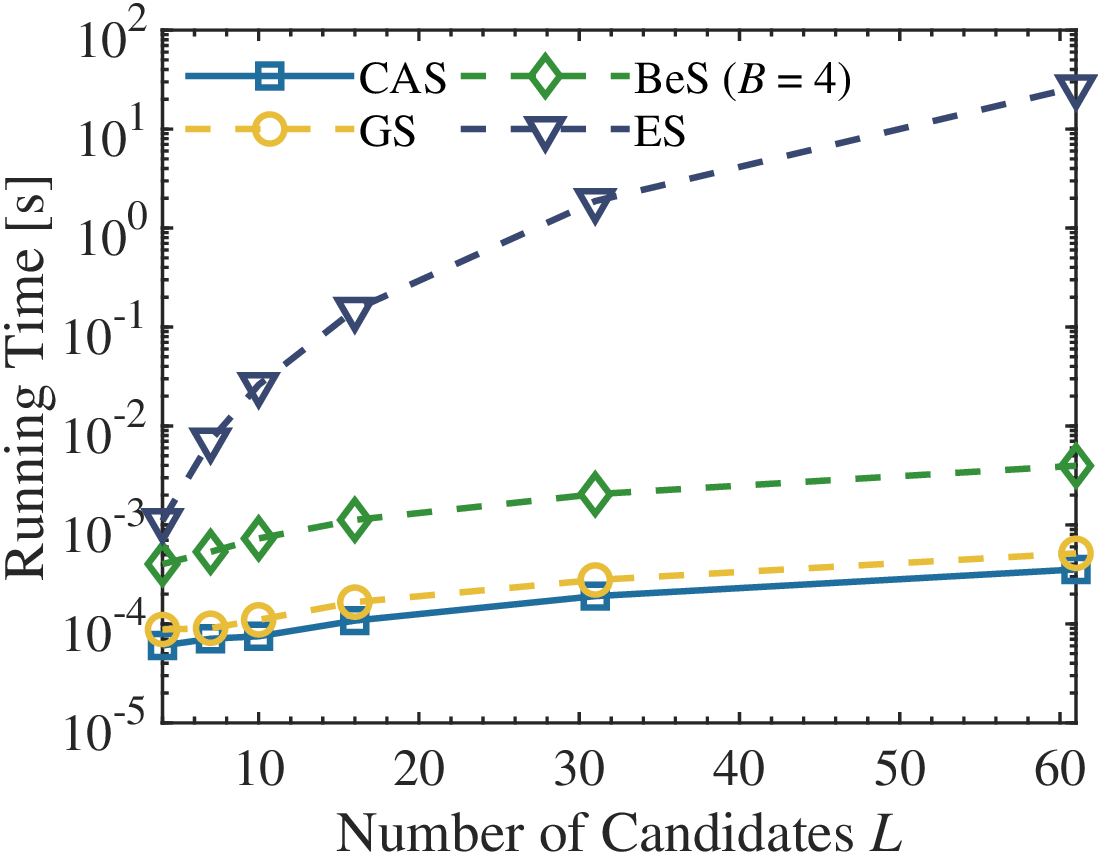}
	   \label{fig:fig3b_runtime_L}
    }
\caption{Impact of the number of candidates.}
\label{fig:fig3a}
\vspace{-10pt}
\end{figure} 

Fig.~\ref{fig:fig_mse_beam_width} evaluates the beam width. CAS and GS are independent of $B$ and serve as constant low-complexity baselines. BeS steadily reduces the MSE as $B$ increases because a wider beam preserves more candidate paths for later layers. A moderate beam width is already sufficient to approach ES, so only a small portion of the full tree is usually needed to capture most of the activation gain. The running-time curves are consistent with the complexity analysis in Section IV-D: CAS is the fastest due to its separable metric, GS visits only $ML$ children, BeS grows with $B$ according to $L+(M-1)BL$, and ES has the highest runtime due to its exponential search over $\sum_{m=1}^{M}L^m$ tree nodes.

Fig.~\ref{fig:fig3a} studies the impact of the number of candidate locations. A larger $L$ reduces the candidate spacing and gives each PA more freedom to match the user geometry, so all PASS-based methods achieve lower MSE. This confirms the value of discrete activation: a finite set of practical activation points can still provide substantial spatial flexibility for AirComp. CAS remains attractive because its search cost scales linearly with $ML$. GS and BeS further reduce the MSE through recursive updates, and BeS stays close to ES with a suitable beam width. Fig.~\ref{fig:fig3b_runtime_L} shows that the runtime grows mildly for CAS and GS, more visibly for BeS, and most significantly for ES. Thus, CAS, GS, and BeS provide practical fast alternatives, while ES serves mainly as an exact benchmark.

\section{Conclusion}
We investigated fast discrete PA activation for AirComp in an uplink multiuser PASS. We derived the optimized AirComp MSE and an exact rank-one recursion for candidate evaluation. We then developed GS and BeS from the search tree and proposed CAS as a separable low-complexity rule with low-SNR optimality. The numerical results showed that discrete PASS activation substantially improved AirComp accuracy over a fixed array, while the proposed methods offered practical complexity-performance tradeoffs.

\clearpage
\bibliographystyle{IEEEtran}
\bibliography{mybib}
\end{document}